\UseRawInputEncoding 
\documentclass[preprint,superscriptaddress,amsmath,amssymb,prd,aps,showpacs,floatfix,nofootinbib]{revtex4-1}
\usepackage{graphicx}
\usepackage{dcolumn}
\usepackage{bm}
\usepackage{mathrsfs}
\usepackage{hyperref}
\usepackage{amsmath}
\usepackage{amssymb}
\usepackage{bm}
\usepackage{color}
\usepackage{float}
\usepackage{dcolumn}
\usepackage{multirow}
\usepackage{changepage}
\usepackage{enumerate}
\hypersetup{pdftex,colorlinks=true,linkcolor=blue,citecolor=red,menucolor=black,urlcolor=blue,filecolor=blue}

\newcommand{\bl}[1]{\textcolor{blue}{#1}}
\begin{document}
\title{Prediction of new $T_{cc}$ states of $D^{*} D^{*}$ and $D^{*}_s D^{*}$ molecular nature}
\date{\today}

\author{L.~R. Dai}
\email{dailianrong@zjhu.edu.cn}
\affiliation{School of Science, Huzhou University, Huzhou 313000, Zhejiang, China}
\affiliation{Departamento de F\'{\i}sica Te\'orica and IFIC, Centro Mixto Universidad de Valencia-CSIC Institutos de Investigaci\'on de Paterna, Aptdo.22085, 46071 Valencia, Spain}

\author{R.~Molina}
\email{Raquel.Molina@ific.uv.es}
\affiliation{Departamento de F\'{\i}sica Te\'orica and IFIC, Centro Mixto Universidad de Valencia-CSIC Institutos de Investigaci\'on de Paterna, Aptdo.22085, 46071 Valencia, Spain}

\author{E.~Oset}
\email{oset@ific.uv.es}
\affiliation{Departamento de F\'{\i}sica Te\'orica and IFIC, Centro Mixto Universidad de
Valencia-CSIC Institutos de Investigaci\'on de Paterna, Aptdo.22085,
46071 Valencia, Spain}

\begin{abstract}
We extend the theoretical framework used of describe the $T_{cc}$ state as a molecular state of $D^* D$ and make predictions for the $D^*D^*$ and $D^*_s D^*$ systems, finding that they lead to bound states only in the $J^P=1^+$ channel. Using input needed to describe the $T_{cc}$ state, basically one parameter to regularize the loops of the Bethe-Salpeter equation, we find bound states with bindings of the order of the MeV and similar widths
 for $D^*D^*$ system, while the $D^{*}_s D^{*}$ system develops a strong cusp around threshold.
 \end{abstract}

\maketitle


\section{Introduction}  The discovery of the mesonic $T_{cc}$ state with two open charm quarks \cite{1f,2f,3f,4f} has brought a new example of an exotic meson state, challenging the standard nature of mesons as $q \bar q$ states. The reaction of the theory community has been fast and many articles have been written proposing different interpretations about the nature of the state \cite{9a,10a,11a,12a,13a,14a,15a,16a,17a,18a,19a,21a,22a,23a,24a,25a,26a,27a,28a,29a,30a,dong,abreu}.
Rates in production reactions had been calculated prior to the $T_{cc}$ discovery \cite{qq}.
The relevant analysis carried out in \cite{6a} by means of a unitary amplitude, with explicit consideration of the  experimental resolution, has shown that the width of the $T_{cc}$ state is much smaller than the one extracted from the raw data in \cite{4f}, and in line with the width obtained in \cite{12a}. Similar conclusions are obtained in \cite{miguel}. The proximity of the the peak position in \cite{6a} to the $D^* D$ threshold strongly supports the $D^* D$ nature of this state, something assumed in most of the works done on the $T_{cc}$ state. In \cite{12a} the state was studied within a unitary coupled channel approach with the $D^{*+} D^0$ and $D^{*0} D^+$ channels and the interaction was obtained from the exchange of vector mesons in a straight extrapolation of the local hidden gauge approach \cite{hidden1,hidden2,hidden4,hideko} to the charm sector \cite{wuraquel}. The only parameter in \cite{12a} was a regulator in the meson meson loop function of the Bethe Salpeter equation, which was tuned to obtain the mass of the state at the right position, then the width and the $D^0 D^0 \pi^+$ mass distribution were obtained in good agreement with experiment, with a width of around $43$ KeV.

   Given the fact that heavy quark spin symmetry \cite{isgur,neubert,manohar} allows to relate the $D$ and $D^*$ sectors, it is tempting to extend the results of \cite{12a} to the $D^* D^*$ system to make predictions on this state. The task is rendered easier because this system and the $D^*_s D^*$ were studied before the recent experimental finding on the $D^* D$ system \cite{4f} in \cite{tania}.  Indeed, in \cite{tania} it was found that the $D^* D^*$ system in isospin $I=0$, and $J^P=1^+$
and the $D^*_s D^*$ system in $ I=\frac{1}{2}, J^P=1^+$ had an attractive potential,  strong enough to support a bound state, and predictions were done with a binding of around $35$ MeV. These predictions are tied to the regulator of the meson meson loop function, and right now we have experimental information from \cite{4f,6a} to fix it, such that more accurate predictions can be done right now. On the other hand, in \cite{tania} the width of the states was obtained from the pseudoscalar-pseudoscalar decay channel, which in the present case is forbidden by spin-parity conservation. Yet, there is another source of decay which is the pseudoscalar-vector channel, which we will evaluate here. This decay channel was investigated in the study of the $D^* \bar{K}^*$ system \cite{raqueldsks} which produced a bound state in \cite{tania} and was shown in  \cite{raqueldsks} to be suited to reproduce the properties of the $X_0(2866)$ state recently observed by the LHCb collaboration in \cite{x0expe}.

  The width of the $T_{cc}$ state is tied to the $D^* \to \pi D$ decay \cite{6a}, and results in about $40-50$ KeV. On the contrary, here the $D^* D^*$ and $D^*_s D^*$ systems will decay into a pseudoscalar-vector system which has a much larger phase space for decay than the $D^* \to  \pi D$. Hence, we can already guess that we shall have a much larger width, yet, still reasonably small, as one can induce from the results of \cite{raqueldsks}.

    The $D^* D^*$ system has been relatively unexplored. It is studied in \cite{shilin} using meson exchange for the interaction  and in \cite{guozougen} using the exchange of vector mesons guided by the local hidden gauge approach.  Only the diagonal interaction and no decay channels are considered in \cite{guozougen}, which aims at providing guidelines on possible states at the qualitative level. The $D^* D^*$ system in $I=0$ and $J^P=1^+$ appears in \cite{guozougen} as a candidate for a bound state, and in \cite{miguel,hanhart} it also appears as a bound state using arguments  of heavy quark spin symmetry and the data of \cite{4f}.
    In \cite{shilin} a state of $D^{(*)} D^{(*)}$ nature, without distinguishing between
$D^* D$ and $D^* D^*$, is also found to be a good candidate for a bound state in $I=0$ and $J^P=1^+$.
Similarly, a possible $(D^{(*)}D^{(*)})_s$ state in $J^P=1^+$ is also reported in \cite{shilin}. The widths are not considered in this latter work.

   With that background and the valuable information from the $T_{cc}$ state, we retake the task of studying the  $D^* D^*$ and $D^*_s D^*$ systems, paying attention to the decay channels, with the purpose of making a precise determination of the mass and widths of the bound states emerging from the interaction of these systems.

\section{FORMALISM}

\subsection{Direct interaction}
The interaction of $D^* D^*$ and $D^*_s D^*$ is studied in \cite{tania}.  It  is  based on the
extrapolation of the local hidden gauge approach \cite{hidden1,hidden2,hidden4,hideko} to the charm sector.
The local hidden gauge approach  was first used in \cite{raquelvec,gengvec} to study the interaction between
vector mesons in the SU(3) sector. It contains a  contact term and  the exchange of vector mesons which requires a
three vector vertex. In \cite{raquelvec,gengvec} this interaction was shown  to produce  bound states or resonances  which could be associated  to existing states.  Extrapolated to  the charm sector, it predicted
the pentaquark states with hidden  charm and hidden   charm and strangeness \cite{wuraquel,wuprc} which were found later by the LHCb collaboration \cite{penta1,penta2,pentastr}.

\begin{figure}[h]
\centering
\includegraphics[scale=0.85]{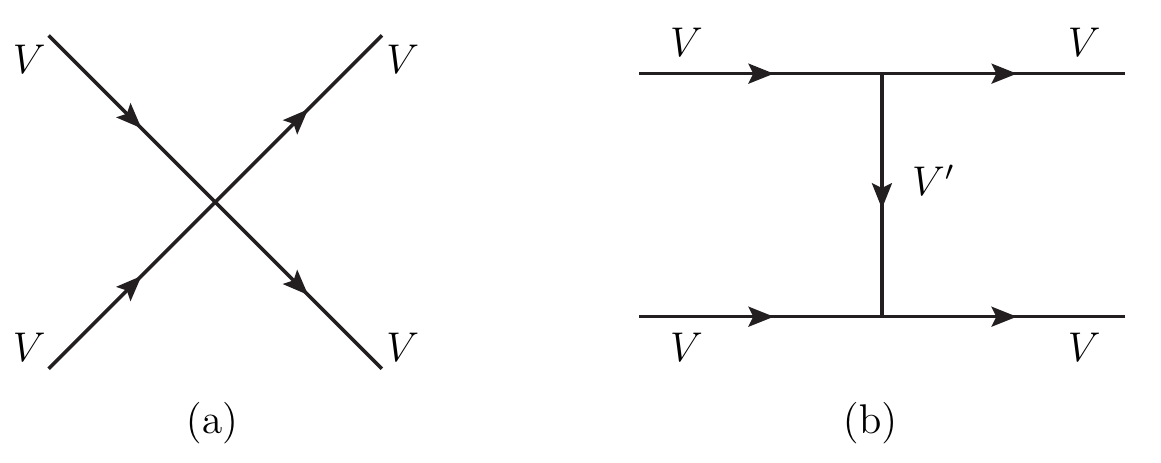}
\caption{Terms for the $VV$ interaction: (a) contact term; (b) vector exchange}
\label{fig:vv}
\end{figure}

  The basic ingredients to calculate the potential between two vectors are the Lagrangians

\begin{eqnarray}
  {\cal{L}}^{(c)} &=& \frac{g^2}{2}\, \langle V_\mu V_\nu V^\mu V^\nu-V_\nu V_\mu V^\mu V^\nu\rangle,
\end{eqnarray}
with $g=\frac{M_V}{2\,f} ~(M_V=800 \, {\rm MeV},~f=93 \, {\rm MeV})$, and  $V_\mu$ the $q\bar{q}$ matrix written in terms of vector mesons
\begin{equation}
V_\mu=\left(
\begin{array}{cccc}
\frac{\omega}{\sqrt{2}}+\frac{\rho^0}{\sqrt{2}} & \rho^+ & K^{*+}&\bar{D}^{*0}\\
\rho^- &\frac{\omega}{\sqrt{2}}-\frac{\rho^0}{\sqrt{2}} & K^{*0}&D^{*-}\\
K^{*-} & \bar{K}^{*0} &\phi&D^{*-}_s\\
D^{*0}&D^{*+}&D^{*+}_s&J/\psi
\end{array}
\right)_\mu\ .
\label{eq:vfields}
\end{equation}
and
\begin{eqnarray}
{\cal{L}}_{VVV} = ig \,\langle V^\mu \partial_\nu V_\mu-\partial_\nu V_\mu V^\mu \rangle.
\end{eqnarray}
${\cal{L}}^{(c)}$ is a contact term and ${\cal{L}}_{VVV}$ stands for the three vector vertex. By means of it, one generates an interaction between
vectors exchanging vector mesons. The mechanisms for the interaction are depicted in Fig.~\ref{fig:vv}.

In table XVI of \cite{tania} it was shown that the contact term for $D^* D^*$ in $I=0$ gives no contribution. By contrast, the vector exchange term
gives null contribution for spin $J=0,2$ but produces an attractive potential in $J^P=1^+$,
\begin{eqnarray}
V_{D^*D^*\to D^*D^*}=\frac{1}{4}g^2(\frac{2}{m_{J/\psi}^2}+\frac{1}{m_\omega^2}-\frac{3}{m_\rho^2}) \lbrace(p_1+p_4).(p_2+p_3)+(p_1+p_3).(p_2+p_4) \rbrace \,.
\label{eq:c1}
\end{eqnarray}
In table XVII of \cite{tania} it is shown that for $I=1$ the interaction for  $J=0,2$ is repulsive and null for $J=1$. The situation is similar for the
$D^*_s D^*_s$ system, which has $I=0$, with repulsive  interaction in  $J=0,2$ and null  interaction for $J=1$ (see table XIX of \cite{tania}).
On the contrary, as shown in table XVIII of \cite{tania}, the $D^*_s D^*$ system, which has
$I=\frac{1}{2}$, gives attraction for $I=1$ and repulsive for $J=0,2$.

 We are thus left with two candidates  for bound states $D^*D^*$ with $I=0,J^P=1^+$ and  $D^*_s D^*$  with $I=\frac{1}{2}, J^P=1^+$. The interaction for this latter case
 is given by
\begin{eqnarray}
V_{D^*_s D^*\to D^*_sD^*}=-\frac{g^2(p_1+p_4).(p_2+p_3)}{m_{K^*}^2}+\frac{g^2(p_1+p_3).(p_2+p_4)}{m_{J/\psi^2}}
\label{eq:c2}
\end{eqnarray}
Note that $(p_1+p_3).(p_2+p_4)$ projected in $s$-wave can be written as \cite{rocaaxial}
\begin{eqnarray}
\frac{1}{2}\lbrace 3s-(M^2_1+M^2_2+M^2_3+M^2_4)-\frac{1}{s}(M^2_1-M^2_2)(M^2_3-M^2_4)\rbrace \,.
\end{eqnarray}
In \cite{tania} the T-matrix was obtained from these potentials using the Bethe-Salpeter equation
\begin{eqnarray}
T=[1-VG]^{-1}V
\label{eq:BS}
\end{eqnarray}
with $G$ the intermediate vector-vector ($VV$) loop function which was regularized by means of a cutoff  and also dimensional regularization.
On the other hand, the pseudoscalar-pseudoscalar ($PP$) decay channels were considered for all the states studied there, but the $VV$ states
with $1^+$ cannot decay to $PP$ if we want to conserve spin and parity. Here we consider instead the decay into vector-pseudoscalar $(VP)$ channel
which will give a width to the bound states that we find.

\subsection{Vector-pseudoscalar decay channels}
\subsubsection{$D^* D^* \to D^* D $ decay}
\begin{figure}[h]
\centering
\includegraphics[scale=0.85]{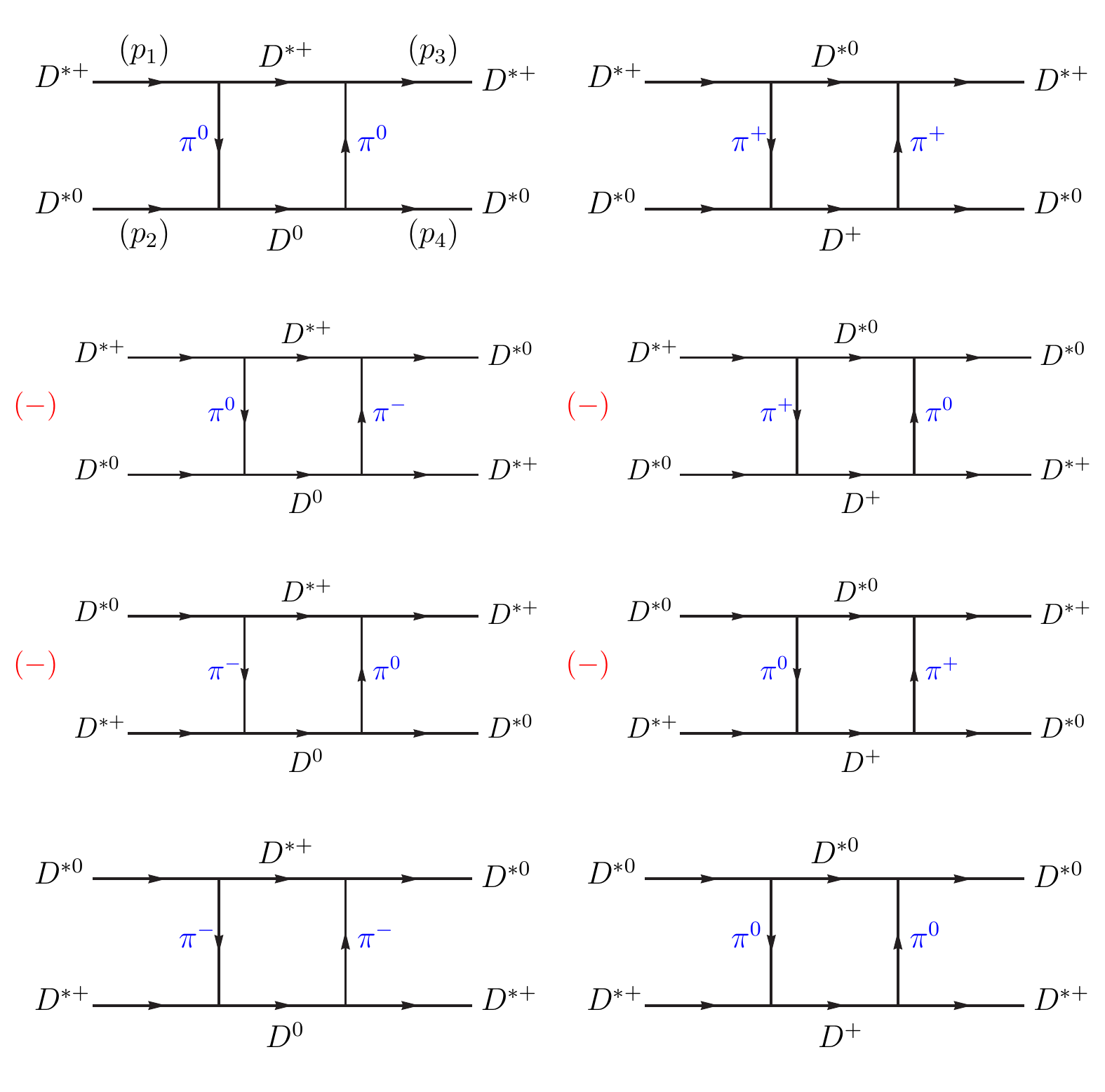}
\caption{Box diagrams according for $D^* D^*,I=0$ decay into $D^{*+}D^{0}$ and $D^{*0}D^{+}$}
\label{fig:box1}
\end{figure}

We take the $I=0$ $D^* D^*$ state. With the isospin doublets $(D^+,-D^0)$ and $(D^{*+},-D^{*0})$, the $I=0$ state is given by
\begin{eqnarray}
|D^* D^*,I=0\rangle=-\frac{1}{\sqrt{2}}|D^{*+}D^{*0}-D^{*0}D^{*+} \rangle \,.
\end{eqnarray}
This system can decay into $D^{*+}D^{0}$ or $D^{*0}D^{+}$ and we shall take  into account these decays by means of the imaginary part of the box diagrams of Fig.~\ref{fig:box1}.
The diagrams shown in  Fig.~\ref{fig:box1} have all the same structure and only the isospin coefficients are different. Taking the first diagram as reference and the coupling
of $\pi^0$ to $D^{*+} D^{*+}$  as $1$, by means of  Clebsch-Gordan coefficients we find the weight  $-1$ for $\pi^0 D^{*0} D^{*0}$, and $\sqrt{2}$ for $\pi^+$ or $\pi^-$ coupling to
$D^{*+} D^{*0}$ (consistent with our phase convention $\pi^+=-|11\rangle$). The total weight of the diagrams is
$$ \frac{1}{4}(1+2+2+4+4+2+2+1)=\frac{18}{4}=\frac{9}{2}$$

\begin{figure}[h]
\centering
\includegraphics[scale=0.85]{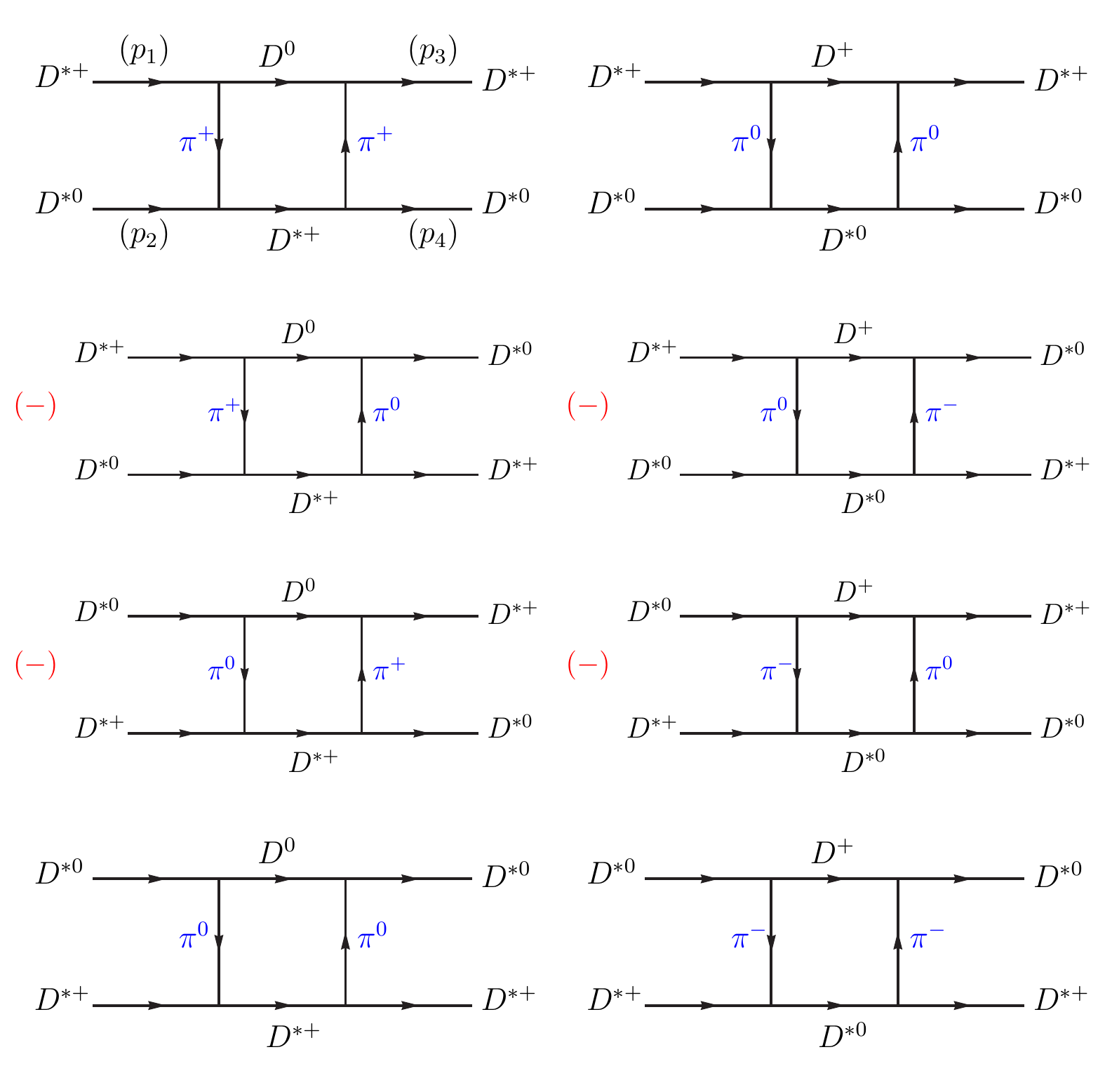}
\caption{Box diagrams according for the $D^* D^*,I=0$ decay into $D^{0}D^{*+}$ and $D^{+}D^{*0}$}
\label{fig:box2}
\end{figure}

\begin{figure}[h]
\centering
\includegraphics[scale=0.85]{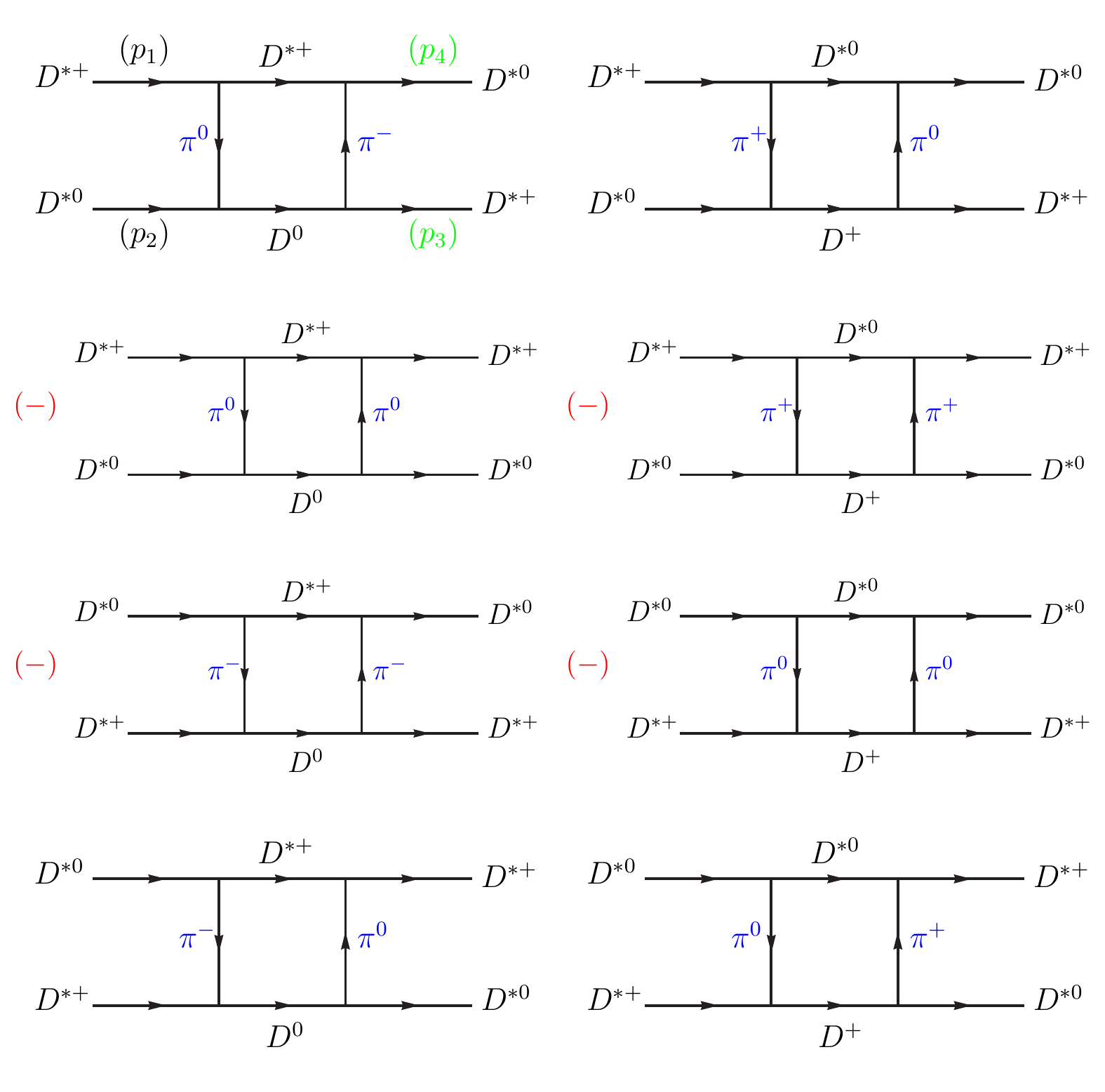}
\caption{Diagrams obtained from Fig.~\ref{fig:box1} exchanges $D^*(p_3) \leftrightarrow D^*(p_4)$  in the final state.}
\label{fig:box3}
\end{figure}

\begin{figure}[h]
\centering
\includegraphics[scale=0.85]{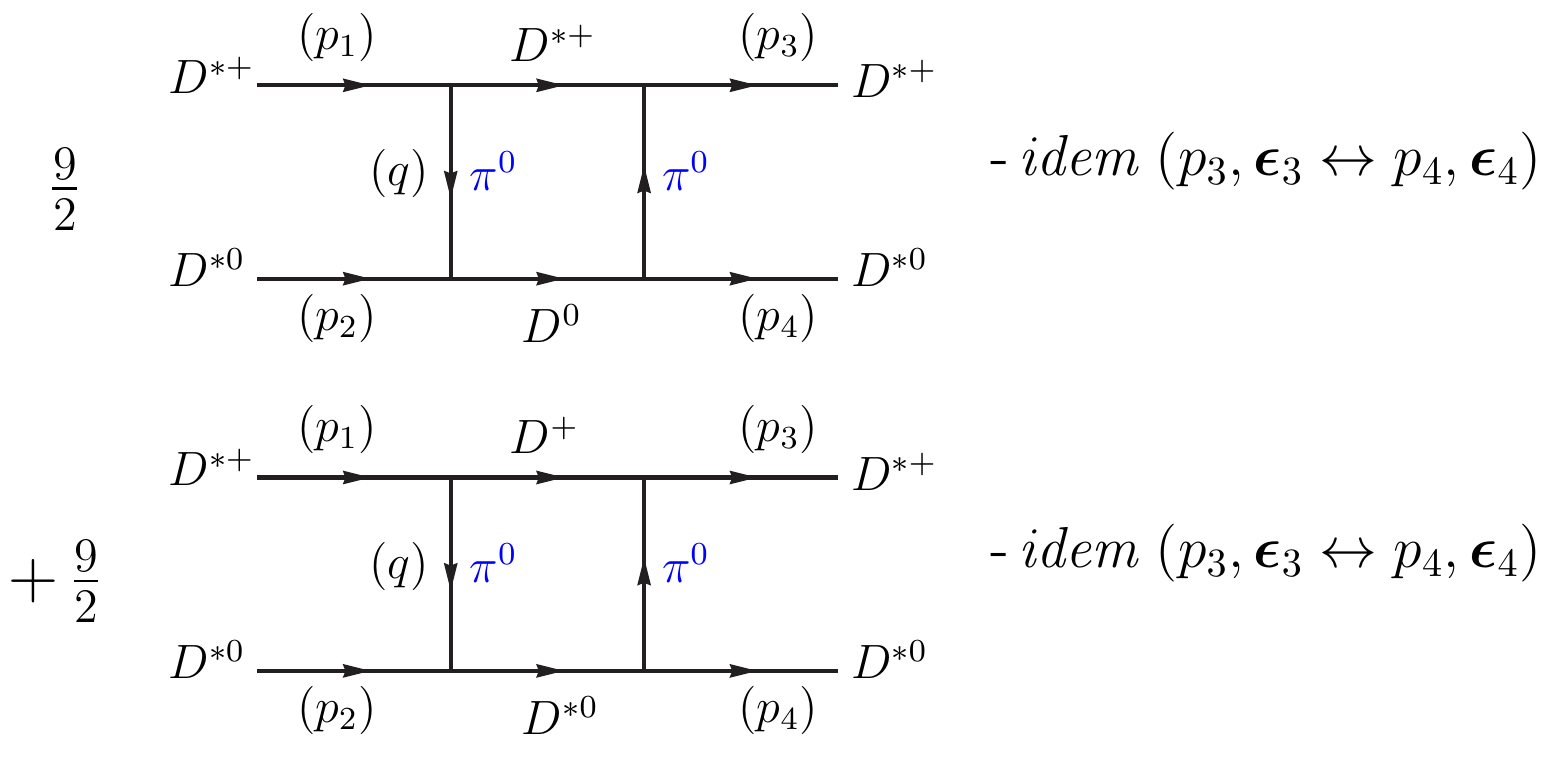}
\caption{Diagrams to be calculated with their respective weights.}
\label{fig:box4}
\end{figure}

We have to consider in addition the diagrams where the pseudoscalar meson is on the upper line of the diagram and the vector in the lower line, which are deplicted in Fig.~\ref{fig:box2}.
If we take the second diagram of Fig.~\ref{fig:box2} as reference (with the exchange of two  $\pi^0$), the rest of them are included as before and altogether we have a weight $\frac{9}{2}$ of
the second diagram of Fig.~\ref{fig:box2}. The set of diagrams must be completed exchanging the vectors $D^*(p_3) \leftrightarrow D^*(p_4)$ in the final state, given the identity of the two
$D^*$ in the final state (in the isospin formalism). Then the diagrams of Fig.~\ref{fig:box1} give rise to the diagrams of Fig.~\ref{fig:box3}.
We observe now that the third diagram of Fig.~\ref{fig:box3} (the one with two $\pi^0$ exchange) is equivalent to the first diagram of Fig.~\ref{fig:box1}, except that $p_3,\bm{\epsilon}_3\leftrightarrow p_4,\bm{\epsilon}_4$
 ($\bm{\epsilon}_i$ is the polarization vector of particle $i$) are exchanged and there is a relative $(-1)$ sign. The same  happens when we exchange the final states of Fig.~\ref{fig:box2},
 which we do not depict. Altogether, the sum of the $32$ diagrams can be calculated as shown in Fig.~\ref{fig:box4}.

The evaluation of these diagrams requires now the use of two new vertices, the ordinary $VPP$ coupling and the anomalous $VVP$ coupling given by the
Lagrangians
\begin{eqnarray}
{\cal{L}}_{VPP} = -ig \,\langle [P, \partial_\mu P] V^\mu \rangle
\end{eqnarray}
\begin{eqnarray}
{\cal{L}}_{VVP} = \frac{G^{\prime}}{\sqrt{2}} \, \epsilon^{\mu\nu\alpha\beta}  \langle  \partial_\mu V_\nu \partial_\alpha V_\beta P \rangle \,.
\end{eqnarray}
${\cal{L}}_{VPP}$ appears in the local hidden gauge approach and ${\cal{L}}_{VVP}$ can be found from \cite{bramon2,pelaezroca},
where $G^{\prime}$ is given by
$$
G^{\prime}=\frac{3\,g^{\prime}}{4\pi^2 f};\qquad g^{\prime}=-\frac{G_V m_\rho}{\sqrt{2} f^2};\qquad G_V=55\, {\rm MeV};\qquad f=93\, {\rm MeV}
$$
In addition to $V^\mu$ of Eq.~\eqref{eq:vfields} we need now the matrix $P$ for the pseudoscalar mesons given by
\begin{equation}
P=\left(
\begin{array}{cccc}
\frac{\eta}{\sqrt{3}}+\frac{\eta'}{\sqrt{6}}+\frac{\pi^0}{\sqrt{2}} & \pi^+ & K^+&\bar{D}^0\\
\pi^- &\frac{\eta}{\sqrt{3}}+\frac{\eta'}{\sqrt{6}}-\frac{\pi^0}{\sqrt{2}} & K^{0}&D^-\\
K^{-} & \bar{K}^{0} &-\frac{\eta}{\sqrt{3}}+\sqrt{\frac{2}{3}}\eta'&D^-_s\\
D^0&D^+&D^+_s&\eta_c
\end{array}
\right)
\label{eq:pfields}
\end{equation}
where the standard $\eta,\eta^{\prime}$ mixing of \cite{bramon} has been used.

  We evaluate the amplitudes at the $D^* D^*$ threshhold since we expect small binding energies. We have then  $\epsilon^0=0$ for all
  the external vector mesons and we take it also for the propagating vectors in the loop given the large mass of the particles. We
  get the following vertices:
\begin{enumerate}[1)]
  \item  $D^{*0} \pi^{0} \to D^0$,  \\
$-i\, t=-2 i\, g \,{\bm q} \, {\bm \epsilon} (D^{*0}) \frac{1}{\sqrt{2}}$
   \item $D^0 \to \pi^{0}  D^{*0}$,\\
 $-i \,t=-2 i\, g \,{\bm q} \, {\bm \epsilon} (D^{*0}) \frac{1}{\sqrt{2}}$
 \item $D^{*+} \to \pi^{0}  D^{*+}$,\\
 $-i \,t= -i \, \frac{G^{\prime}}{\sqrt{2}} \, \epsilon^{ijk} E(D^{*+}_{\rm ext}) \,\epsilon_{i}(D^{*+}_{\rm ext}) \, q_j \,  \epsilon_{k}({\rm int}) \,\frac{1}{\sqrt{2}}$
   \item $D^{*+} \pi^{0} \to D^{*+} $,\\
 $-i \,t= -i \, \frac{G^{\prime}}{\sqrt{2}} \, \epsilon^{ijk} E(D^{*+}_{\rm ext}) \,  \epsilon_{i}(D^{*+}_{\rm ext}) \, q_j \,  \epsilon_{k}({\rm int}) \,\frac{1}{\sqrt{2}}$
\end{enumerate}
where all vector and tensor components are contravariant (even if we write them as lower indices) and $E(D^*)$ stands for the energy of the $D^*$. The indices  ext or int stand for the external or internal vectors of the
diagrams.

Taking into account that
$$
\sum_{pole} \epsilon_k(\rm int)\epsilon_{k^{\prime}}(\rm int) =\delta_{k k^{\prime}}
$$
the product of all four vertices gives
\begin{eqnarray}
(\sqrt{2}g)^2\big(\frac{G'}{2}\big)^2 E(1)E(3)\lbrace \epsilon_i(1)\epsilon_l(2)\epsilon_i(3)\epsilon_m(4){\bm q}^2 q_l q_m-\epsilon_j(1)\epsilon_l(2)\epsilon_i(3)\epsilon_m(4)q_i q_j q_l q_m \rbrace
 \label{eq:8p1}
 \end{eqnarray}
where the indices $1,2,3,4$ refer to the particles on the order of Fig.~\ref{fig:box4}. Let us note that at threshold all the propagators in the loop depend only on ${\bm q}^2$ which allows us to write
\begin{eqnarray}
 \int d^3q f({\bm q}\,^2) q_l q_m &=&  \int d^3q f({\bm q}\,^2)   \frac{1}{3} {\bm q}^2  \delta_{lm} \nonumber\\
 \int d^3q f({\bm q}\,^2)q_i q_j q_l q_m &=&  \int d^3q f({\bm q}\,^2)   \frac{1}{15} {\bm q}^4  (\delta_{ij}\delta_{lm}+\delta_{il}\delta_{jm}+\delta_{im}\delta_{jl})
\end{eqnarray}
The second combination in Eq.~\eqref{eq:8p1} gives rise to the product of polarization vectors in the order of $1,2,3,4$
\begin{eqnarray}
\epsilon_j \epsilon_l \epsilon_j \epsilon_l+\epsilon_j \epsilon_i \epsilon_i \epsilon_j+\epsilon_j \epsilon_j \epsilon_i \epsilon_i \,,
\label{eq:8p2}
\end{eqnarray}
and using the projectors into the spin states of $J=1,2,3$, ${\cal P}^{(0)}$,${\cal P}^{(1)}$,${\cal P}^{(2)}$ from \cite{raquelvec,raqueldsks} we have
\begin{eqnarray}
\epsilon_j  \epsilon_j \epsilon_i \epsilon_i  &=& 3 {\cal P}^{(0)}  \nonumber\\
\epsilon_j  \epsilon_l \epsilon_j \epsilon_l &= & {\cal P}^{(0)}+{\cal P}^{(1)}+{\cal P}^{(2)}\nonumber\\
\epsilon_j  \epsilon_i \epsilon_i \epsilon_j &= & {\cal P}^{(0)}-{\cal P}^{(1)}+{\cal P}^{(2)} \,.
\label{eq:projmu}
\end{eqnarray}
Hence, the combination of Eq.~\eqref{eq:8p2} give
\begin{eqnarray}
 {\cal P}^{(0)}+{\cal P}^{(1)}+{\cal P}^{(2)}+{\cal P}^{(0)}-{\cal P}^{(1)}+{\cal P}^{(2)}+3 {\cal P}^{(0)}\nonumber
\end{eqnarray}
and we see that this term does not contribute to our state with $J^P=1^+$. The first term of Eq.~\eqref{eq:8p1} gives rise to the combination
\begin{eqnarray}
\epsilon_i  \epsilon_l \epsilon_i \epsilon_l =  {\cal P}^{(0)}+{\cal P}^{(1)}+{\cal P}^{(2)}  \nonumber
\end{eqnarray}
One can see that the diagrams of Fig.~\ref{fig:box2} give rise to the same combination, and those of Fig.~\ref{fig:box4} and the equivalent to Fig.~\ref{fig:box2} exchange
 $p_3,\bm{\epsilon}_3 \leftrightarrow p_4,\bm{\epsilon}_4$ give the same contribution except for a minus sign and the exchange of   $\bm{\epsilon}_3 \leftrightarrow \bm{\epsilon}_4$.
 Hence we get the combination now of
 \begin{eqnarray}
   -\epsilon_i  \epsilon_l \epsilon_l \epsilon_i=-({\cal P}^{(0)}-{\cal P}^{(1)}+{\cal P}^{(2)}) \nonumber
  \end{eqnarray}
and we see that they give the same contribution  to ${\cal P}^{(1)}$ as the other diagrams. Altogether we  find now for the $J^P=1^+$ state the contribution for the four diagrams  of
Fig.~\ref{fig:box4}, keeping the positive energy part of the propagators of the heavy particles
\begin{eqnarray}
 -i\, t &=& 4 \, \frac{9}{2} \,\frac{1}{3} \int\frac{d^4q}{(2\pi)^4}  \, \frac{1}{2 E_{D^*}({\bm q})}  \,  \frac{i}{p^0_1-q^0-E_{D^*}({\bm q})+i\epsilon }  \,\frac{1}{2 E_{D}({\bm q})}
\nonumber\\
&\times &   \, \frac{i}{p^0_2+q^0-E_{D}({\bm q})+i\epsilon }  \, \frac{i}{q^2-m^2_{\pi}+i\epsilon }  \, \frac{i}{(p_2-p_4+q)^2-m^2_{\pi}+i\epsilon } \, {\bm q}^4
\end{eqnarray}
The pion  propagator cannot be placed on shell and if we are interested in the imaginary part only the $D^* D$  intermediate particles can be put on shell. We can
perform the $q^0$ analytically and then use $Im \frac{1}{x+i\epsilon}=-i \pi \delta(x)$ (alternatively one can use Cutkowsky rules) and we find
\begin{eqnarray}
Im V_{\rm box}=-6 \frac{1}{8\pi}\frac{1}{\sqrt{s}} q^5 E^2_{D^*} (\sqrt{2}g)^2 \big(\frac{G'}{2}\big)^2 \left(\frac{1}{(p^0_2-E_{D}({\bm q}))^2-{\bm q}^2-m^2_{\pi}}\right)^2 \,F^4(q)\, F_{HQ}
\label{eq:box1}
\end{eqnarray}
where $$q=\frac{\lambda^{1/2}(s,m^2_{D^*},m^2_{D})}{2\sqrt{s}}\,; \qquad E_{D^*}=\frac{\sqrt{s}}{2}$$
where we have added the form factor $F(q)$ used in \cite{tania,raqueldsks} and the heavy quark correcting factor $F_{HQ}$ to correct the $VPP$ vertex for heavy particles, as discussed in \cite{xiaoliang}
\begin{eqnarray}
F(q)=e^{((q^{0})^2-{\bm q}^2)/\Lambda^2}
\label{eq:form}
\end{eqnarray}
with $$q^0=p^0_1- E_{D^*}({\bm q})$$ and $$F_{HQ}=\left(\frac{m_{D^*}}{m_{K^*}}\right)^2$$

\subsubsection{$D_s^* D^* \to D_s^* D+D_s D^* $ decay}
\begin{figure}[h]
\centering
\includegraphics[scale=0.85]{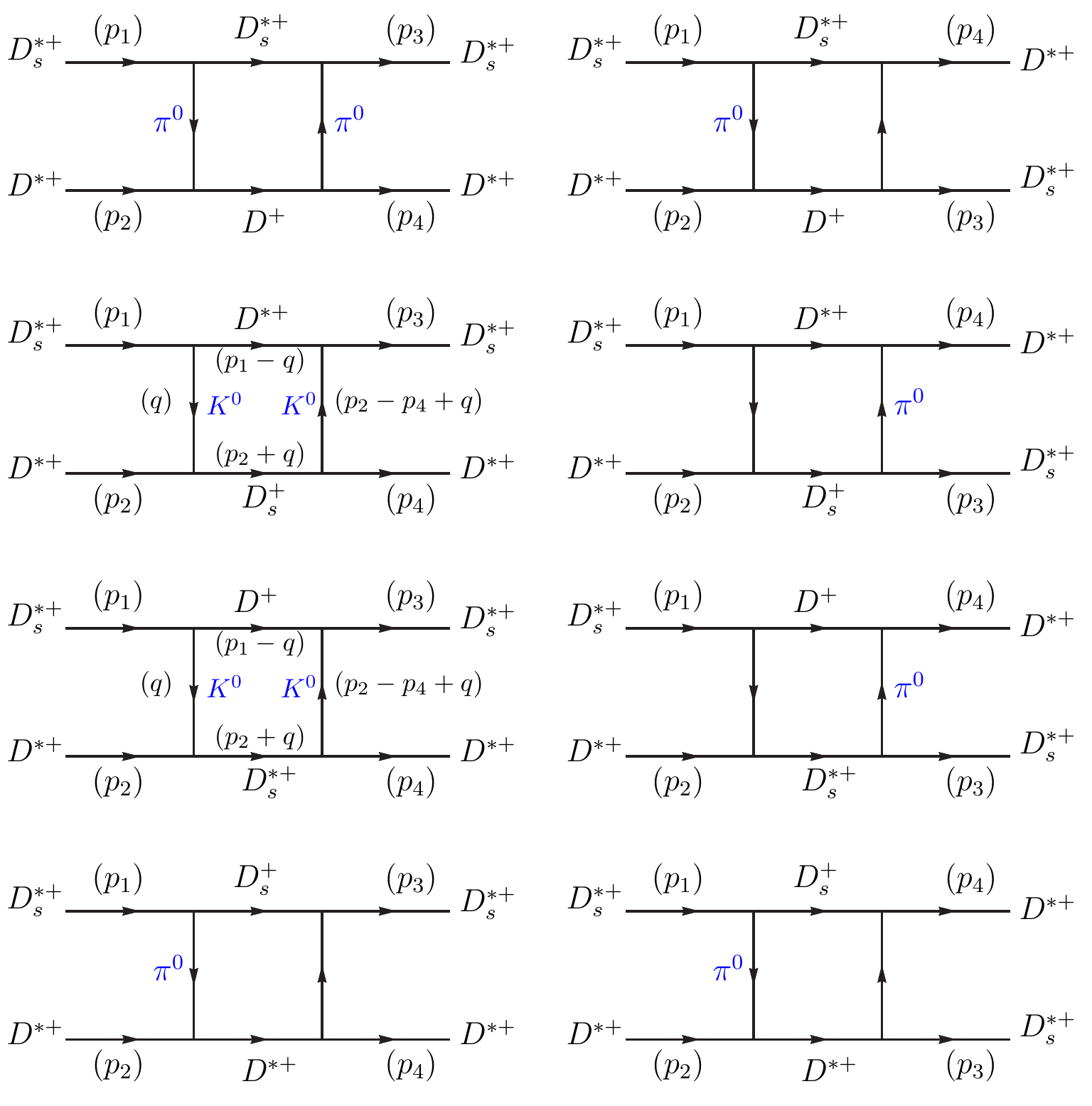}
\caption{Diagrams for the decay of $D_s^{*+}D_s^{*+}$ into $D_s^{*+} D^{+}$ and $D_s^{+} D^{*+}$.}
\label{fig:box5}
\end{figure}

We take the $D_s^{*+} D^{*+}$ state and consider the decay into $D_s^{*+} D^{+}$ and $D_s^{+} D^{*+}$. The diagrams that we must consider are now depicted in Fig.~\ref{fig:box5}.
Since $D_s^*$ does not couple to $D_s^* \pi^0$  we  see that the first and fourth diagrams to the left in Fig.~\ref{fig:box5} are zero and so are all the diagrams  of the right
in Fig.~\ref{fig:box5}
 We repeat the calculations as done in the former subsection, omitting details and we obtain now
\begin{eqnarray}
Im V_{\rm box}=-\frac{1}{3} \frac{1}{8\pi}\frac{1}{\sqrt{s}} (2g)^2 \big(\frac{G'}{\sqrt{2}}\big)^2 (E_1 E_3+E_2 E_4)q^5 \left(\frac{1}{(p^0_2-E_{D_s}({\bm q}))^2-{\bm q}^2-m^2_{K}}\right)^2F^4(q)F_{HQ} ~~~~~~\,
\label{eq:box2}
\end{eqnarray}
with $F(q)$ given by Eq.~\eqref{eq:form} with
\begin{eqnarray}
q^0=p^0_2- E_{D_s}({\bm q})\,;\quad q=\frac{\lambda^{1/2}(s,m^2_{D^*},m^2_{D_s})}{2\sqrt{s}}\,;\quad p^0_2=\frac{s+m^2_{D^*}-m^2_{D^*_s}}{2\sqrt{s}} \nonumber
\end{eqnarray}

 We then solve the  Bethe-Salpeter equation of Eq.~\eqref{eq:BS} with
 \begin{eqnarray}
 V \to V + i\, Im V_{\rm box}   \nonumber
\end{eqnarray}
for  the two cases  $D^* D^*$, $I=0$, and $J^P=1^+$ and $D^*_s D^*$, $ I=\frac{1}{2}, J^P=1^+$ and calculate the T-matrix. By plotting $|T|^2$ we find the mass of the state and  its width which we report in the next section.

\section{Results}
We will use the cut off method to regularize the loops, something advised in \cite{wuzou}. The value of the cut off, or subtraction constant in dimensional regularization, are normally
the only free parameter in the theory, and this is the case here. One can take reasonable cut offs from $450$ MeV to $750$  MeV and make reasonable predictions, but in the present case we  can use the
cut off that was needed in \cite{12a} to get the experimental binding of the $D^* D$ state and rely upon it. We would then be making use of the findings of
\cite{gengalten,genghqs} encouraging the use of the same cut off to respect rules of heavy quark symmetry, although there are limitations to the use of this symmetry in the case
of two heavy quarks as we have here \cite{a,b,c}.

\begin{figure}[h]
\centering
\includegraphics[scale=0.8]{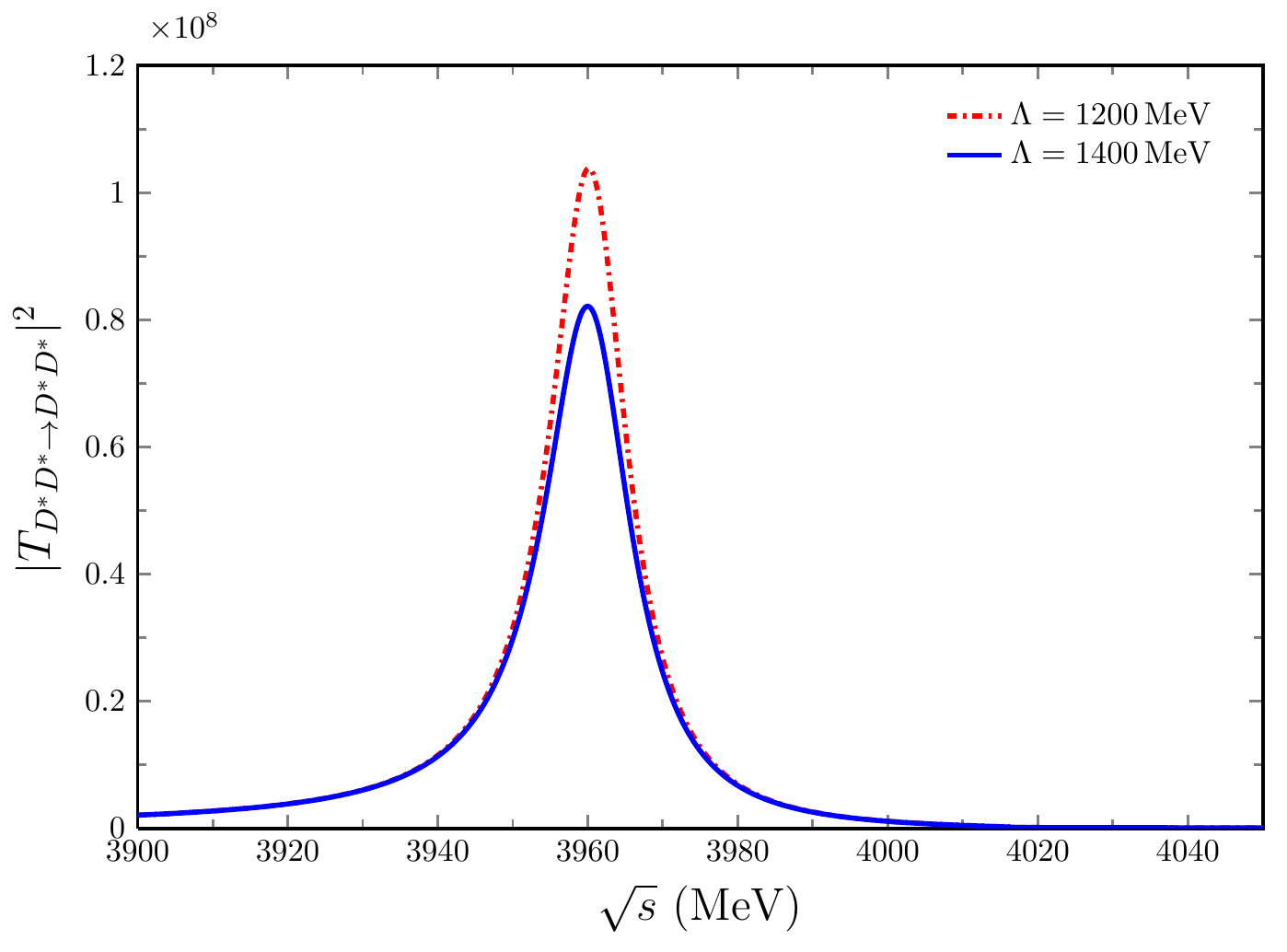}
\caption{Squared amplitude $|T_{D^* D^*\to D^* D^*}|^2 $ with $q_{\rm max}=750 \, {\rm MeV}$.}
\label{fig:c1L}
\end{figure}
\begin{figure}[h]
\centering
\includegraphics[scale=0.8]{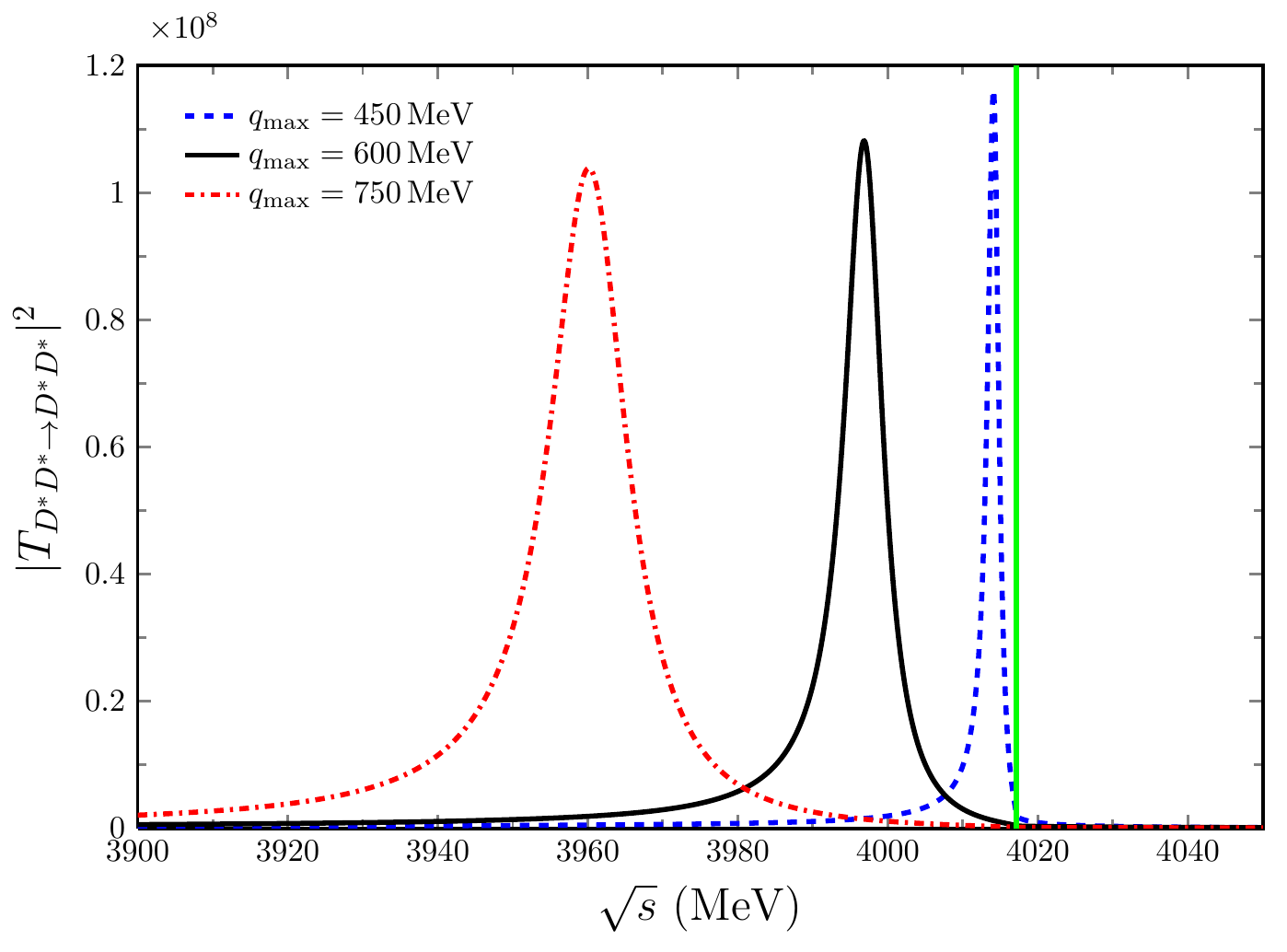}
\caption{Squared amplitude $|T_{D^* D^*\to D^* D^*}|^2 $ with $\Lambda=1200\,{\rm MeV}$. The vertical line indicates the $D^* D^*$ threshold at $4017.1$ MeV. }
\label{fig:c1Q}
\end{figure}

In Fig.~\ref{fig:c1L} we show the results for $|T_{D^*D^*,D^*D^*}|^2$ as a function of $ \sqrt{s}$ calculated with a cut off for $G$ in  Eq.~\eqref{eq:BS},  $q_{\rm max}=750$ MeV, and a value of $\Lambda$ of Eq.~\eqref{eq:form} of $1200$ MeV, similar to what was used in \cite{tania}, and also for $\Lambda =1400$ MeV. We see that the results do not change much by changing the value of $\Lambda$ and we get a bound state around $3960$ MeV.

\begin{figure}[h]
\centering
\includegraphics[scale=0.8]{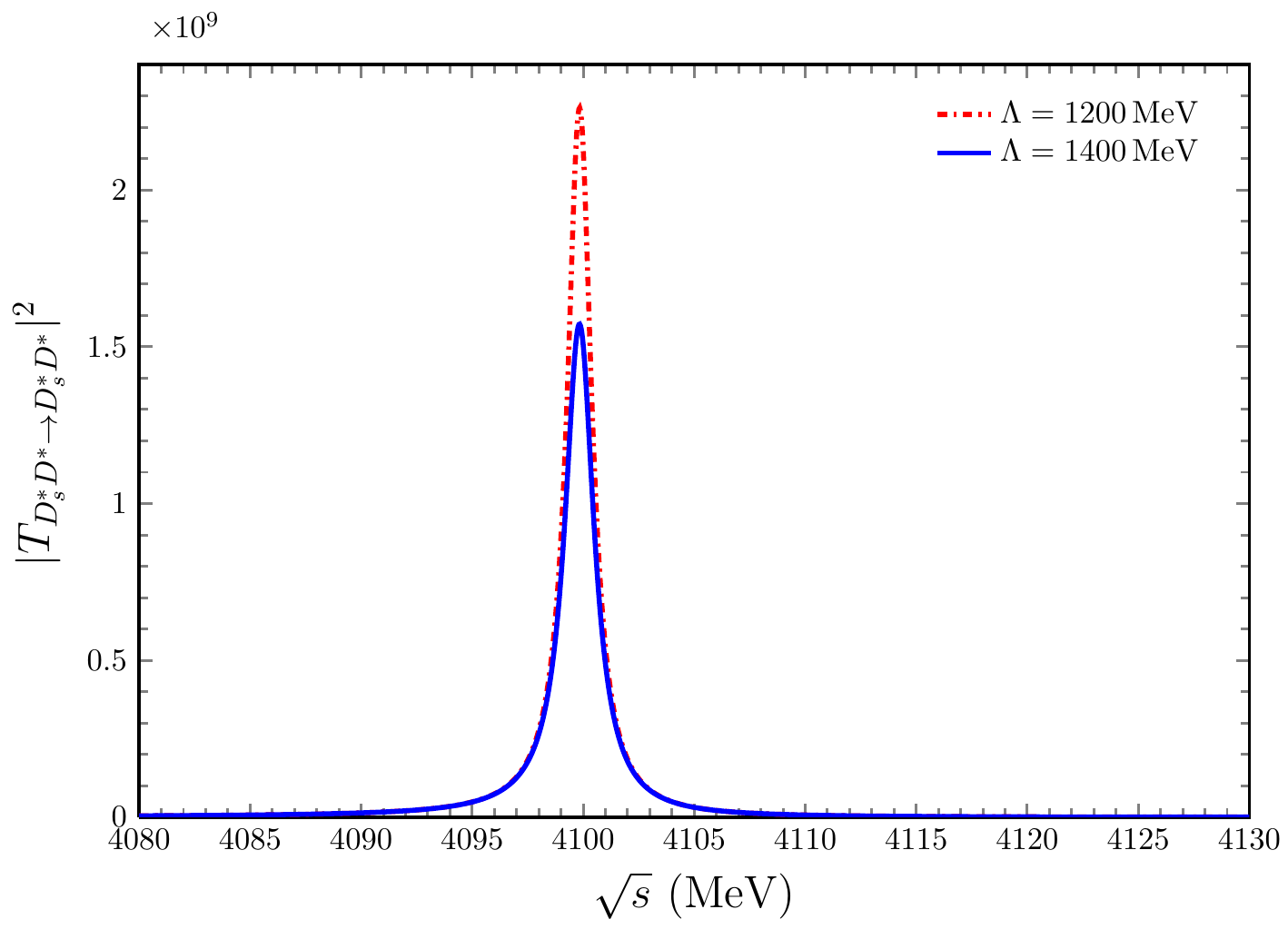}
\caption{Squared amplitude $|T_{D^*_s D^* \to D^*_s D^*}|^2 $ with $q_{\rm max}=750 \, {\rm MeV}$.}
\label{fig:c2L}
\end{figure}
\begin{figure}[h]
\centering
\includegraphics[scale=0.8]{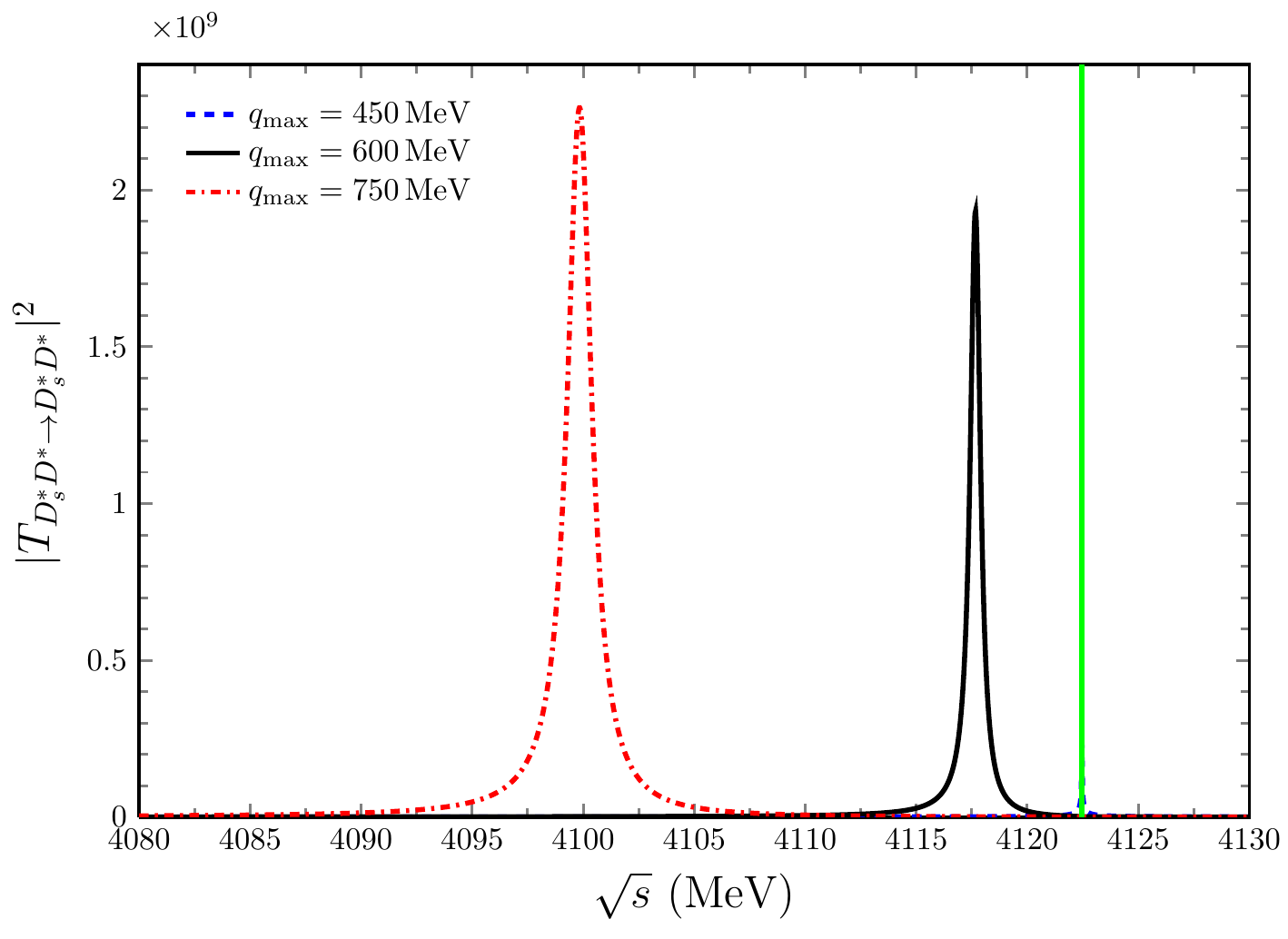}
\caption{Squared amplitude $|T_{D^*_s D^*\to D^*_s D^*}|^2 $ with $\Lambda=1200\,{\rm MeV}$. The vertical line indicates the $D_s^* D^*$ threshold at $4122.46$ MeV.}
\label{fig:c2Q}
\end{figure}

In Fig.~\ref{fig:c1Q} we show instead the results of $|T_{D^*D^*,D^*D^*}|^2$ for a fixed value of $\Lambda=1200$ MeV and different values of the cut off. We can see that there is always a bound state. The binding energy depends on the cutoff value and for values of $q_{\rm max}$ of the order of $450$ MeV, as needed in \cite{12a} to get the $T_{cc}$ state, we also obtain a bound state very close to the $D^*D^*$ threshold. We observe a curious phenomena which is that the width becomes smaller as we get closer to the threshold, in spite of the fact that the phase space for decay increases with increasing energy. To understand this feature we recall that including the box diagram in our approach, adding it to the potential from vector exchange and solving the Bethe-Salpeter equation, is an effective way to include the $D^* D$ channel together with $D^*D^*$, or the $D_s^* D$, $D_s D^*$ channels together with the  $D_s^* D^*$ channel. Then one must recall a well known fact, based on the Weinberg compositeness condition \cite{wein,baru,danijuan}, that the coupling squared of the bound state to the hadron-hadron component in a single channel goes as the square root of the binding energy.  As we go closer to the $D^* D^*$ threshold the coupling of the resonance to this channel becomes smaller. What is less known is that in the case of coupled channels, if we approach one threshold, all the couplings to the different channels that couple to the one of that threshold also go to zero \cite{juantoki,danijuan}. Then the width of the $D^*D^*$ state obtained will be proportional to the square of the coupling of that state to $D^*D$ and will go to zero as we approach the $D^*D^*$ threshold.

In Fig.~\ref{fig:c2L} we show the results for the $D_s^* D^*$ case. We show the results of $|T_{D_s^*D^*,D_s^* D^*}|^2$ for $q_{\rm max}=750$ MeV and two values of the parameter $\Lambda$. As we can see, we get a bound state and the width does not change much with the value of $\Lambda$. In Fig.~\ref{fig:c2Q} we show the results of $|T_{D_s^*D^*,D_s^*D^*}|^2$  for $\Lambda=1200$ MeV and three values of $q_{\rm max}$. We see a similar trend as before, but the bindings are smaller as a consequence of the smaller strength of the potential.

\begin{figure}[h]
\centering
\includegraphics[scale=0.8]{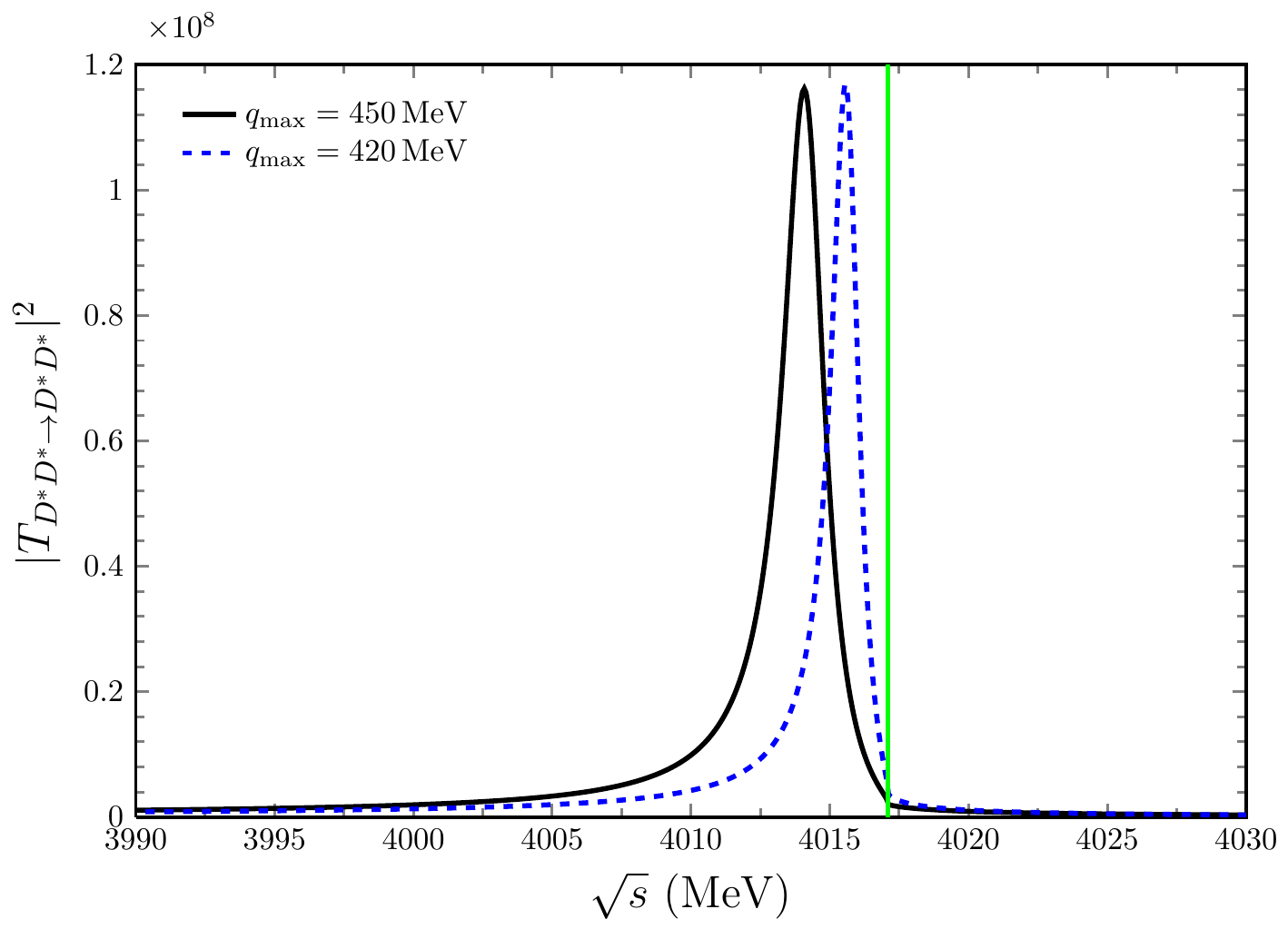}
\caption{The same as Fig.~\ref{fig:c1Q} but with a smaller range of $q_{\rm max}$ as in \cite{12a}.}
\label{fig:c1Qc}
\end{figure}
\begin{figure}[h]
\centering
\includegraphics[scale=0.8]{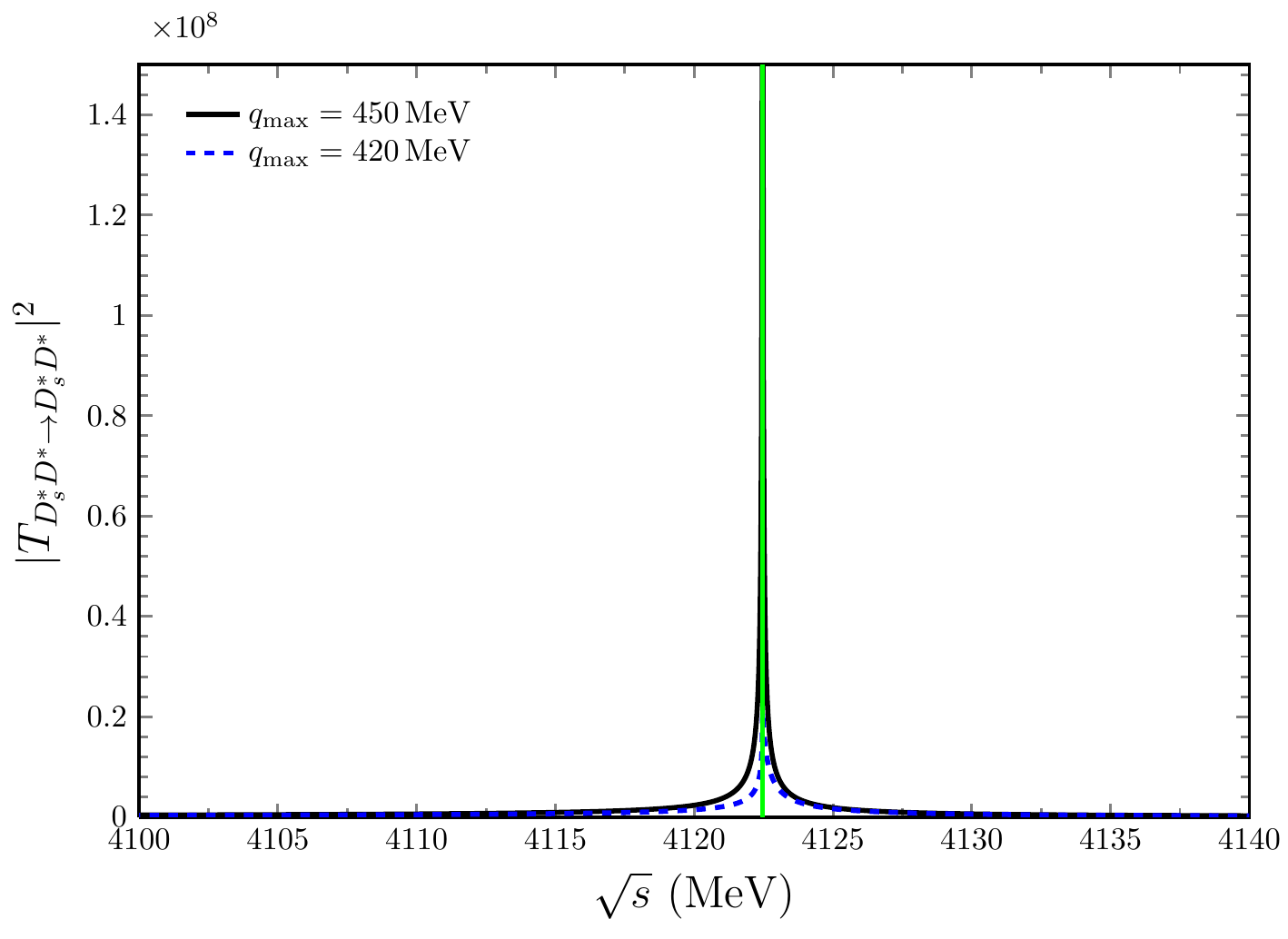}
\caption{The same as Fig.~\ref{fig:c2Q} but with a smaller range of $q_{\rm max}$ as in \cite{12a}.}
\label{fig:c2Qc}
\end{figure}

In Figs.~\ref{fig:c1Qc} and \ref{fig:c2Qc} we show the enlarged picture of the states in Figs.~\ref{fig:c1Q} and \ref{fig:c2Q} for $q_{\rm max}=450$ MeV, together with the results for $q_{\rm max}=420$ MeV (the value taken in \cite{12a} to get the $T_{cc}$ state). For the  $D^{*}_s D^{*}$ system we do not get bound states in these latter cases, and instead we find pronounced cusps at the  $D^{*}_s D^{*}$ threshold.  This is  a consequence
of the weaker potential $V$ in the  case of  $D^{*}_s D^{*}$ compared to $D^{*} D^{*}$, Eqs.~\eqref{eq:c1} and ~\eqref{eq:c2} (see for instance Tables XVI and XVIII of \cite{tania} for numerical values).
 We summarize this information in  Table~\ref{table:MG}, providing the mass and width of the states.
Values of the binding around $0.5-1.5$ MeV for the $D^* D^*$ system are  also obtained in \cite{miguel,hanhart} using arguments of heavy quark spin symmetry and the data of \cite{4f}. The width is not calculated there. The work of \cite{miguel}
 explores the possibility of an $I=1$ state, not fully ruled out by the data, but our theoretical framework excludes such a state. We think these are sensible predictions that should encourage the search of these states at LHCb.

\begin{table}[t]\centering\small%
\caption{The predictions assuming the cut off of the order of $420-450$ MeV as in \cite{12a}. Threshold mass  for $D^* D$, $4017.1$ MeV, and for $D_s^* D^*$, $4122.46$ MeV. }
\begin{tabular}{ccc}%
\hline\hline
&~~~~~  $q_{\rm max}=450~ {\rm MeV}~~~~~$ ~~~~~  &~~~~~ $q_{\rm max}=420~{\rm MeV}$ ~~ \\ \hline
$~~M_{D^*D^*} $        & $4014.08$~MeV    & $4015.54$~MeV \\
$~~B_{D^*D^*} $        & $3.23$~MeV      & $1.56$~MeV \\
$~~\Gamma_{D^*D^*} $        & $2.3$~MeV      & $1.5$~MeV \\
\hline \hline
$~~M_{D_s^*D^*} $        & $4122.46$~MeV~\bl{(cusp)}    & $ 4122.46$ ~MeV ~\bl{(cusp)} \\
$~~\Gamma_{D_s^*D^*}$        &  $ 70-100 $~KeV     &  $ 70-100 $~KeV  \\
\hline \hline
\end{tabular}
\label{table:MG}
\end{table}

\section{Conclusions}
   Encouraged by the recent experimental observation of the $T_{cc}$ state close to the $D^*D$ threshold, that can be explained as a molecular state of $D^*D$, we extend the theoretical work on this latter state to the
   $D^* D^*$ with $I=0$ and $D_s^* D^*$ with $I=1/2$ systems, which have been studied in the past and were shown to develop a bound state with $J^P=1^+$. We also show that these systems with other quantum numbers, or the $D_s^* D_s^*$ system, do not lead to bound states. We have taken advantage of the experimental information on the binding of the $T_{cc}$ state to fix the cutoff regulator of the loops in the Bethe-Salpeter equation. On the other hand, we have included the decay into the $D^*D$ system which involves anomalous couplings, which allows us to get the width of the states, something not done before. The result of our calculations is that both systems give rise to bound states, and assuming the same cut off as was needed in the study of the $T_{cc}$ state, we predict bindings of the order of the MeV and also widths of the same order of magnitude
for the $D^{*} D^{*}$  system, while for the $D^{*}_s D^{*}$ system leads to a pronounced cusp around threshold with a width of the order of $ 70-100 $~KeV. The width of the $D^{*} D^{*}$ system is
 much larger than the one of the $T_{cc}$ state, $40-50$ KeV,  because in this latter case the width is due to the decay of the $D^*$ into $D \pi$,
 for which there is very little phase space, but in the present case we have the decay channel $D^* D$ and there is a much larger phase space for the decay.
 For the  $D^{*}_s D^{*}$ the width is much smaller than for the $D^{*} D^{*}$ state, as consequence of different factors in the formulas of $Im V_{\rm box}$ in Eqs.~\eqref{eq:box1},~\eqref{eq:box2} and the
 fact that one has  $\pi$ exchange in the $D^{*} D^{*}$  case while there is kaon exchange in the case of $D^{*}_s D^{*}$.
  We think the predictions are rather reliable and encourage the LHCb collaboration to look for these states in the near future.

\section*{ACKNOWLEDGEMENT}
This work is partly supported by the National Natural Science Foundation of China under Grants No. 11975009 and No. 12175066.
RM acknowledges support from the CIDEGENT program with Ref. CIDEGENT/2019/015 and from the spanish national grants PID2019-106080GB-C21 and PID2020-112777GB-I00.
This work is also partly supported by the Spanish Ministerio de Economia y Competitividad (MINECO) and European FEDER funds
under Contracts No. FIS2017-84038-C2-1-P B, PID2020-112777GB-I00, and by Generalitat Valenciana under contract PROMETEO/2020/023.
This project has received funding from the European Union Horizon 2020 research and innovation programme under
the program H2020-INFRAIA-2018-1, grant agreement No. 824093 of the ¡°STRONG-2020¡± project.


\end{document}